\documentclass[11pt]{article}
\usepackage{jcappub}
\usepackage{mathtools}
\usepackage{bm}
\usepackage{dsfont}
\usepackage{color}
\usepackage{array}
\usepackage{graphicx}
\usepackage{soul}
\usepackage{multirow}
\usepackage{multicol}
\usepackage{float}
\usepackage{hhline}
\usepackage[dvipsnames]{xcolor}
\usepackage[normalem]{ulem}
\usepackage{mathrsfs}
\usepackage{wasysym}
\usepackage[mathscr]{euscript}
\usepackage{braket}

\title{Dark matter spikes with strongly self-interacting particles}

\author[a,b]{Boris Betancourt Kamenetskaia,}
\author[a,c]{Motoko Fujiwara,}
\author[a]{Alejandro Ibarra,}
\author[d,e]{Takashi Toma}
\affiliation[a]{Technical University of Munich, TUM School of Natural Sciences, Physics Department, 
James-Franck-Str. 1, 85748 Garching, Germany}
\affiliation[b]{\normalsize Max-Planck-Institut f\"ur Physik (Werner-Heisenberg-Institut), Boltzmannstra\ss e 8, 85748 Garching, Germany}
\affiliation[c]{University of Toyama, Department of Physics, 3190 Gofuku, Toyama 930-8555, Japan}
\affiliation[d]{Institute for Theoretical Physics, Kanazawa University, Kanazawa 920-1192, Japan}
\affiliation[e]{Institute of Liberal Arts and Science, Kanazawa University, Kanazawa 920-1192, Japan}

\emailAdd{boris.betancourt@tum.de}
\emailAdd{motoko@sci.u-toyama.ac.jp}
\emailAdd{ibarra@tum.de}
\emailAdd{toma@staff.kanazawa-u.ac.jp}

\abstract
{
    An unavoidable prediction of scenarios with Dark Matter (DM) self-interactions is the existence of number changing processes that convert $n$ initial DM particles into $m$ final ones ($n\to m$ processes), possibly accompanied by Standard Model particles.
    We argue that the $n\rightarrow m$ processes could be probed in DM spikes at the center of galaxies, where the high density may allow sizable rates.
    We systematically study the implications of the $n  \to  m$ processes in DM spikes, including other possible interactions involving DM, such as annihilation and self-scattering.
    We find that for $n\geq3$, the spike is significantly depleted for $n\to m$ cross-sections favored by DM production via thermal freeze-out.
    On the other hand, the semi-annihilation of two DM particles into one DM particle and one Standard Model particle preserves in general the structure of the spike.
    Such density modifications significantly affect phenomenological studies of both astrophysics and particle DM processes around DM spikes.
}

\begin{document}
\begin{flushright}
{\tt 
UT-HET-144
\\
KANAZAWA-25-02
}
\end{flushright}
	\maketitle
	\flushbottom
	
	\section{Introduction} \label{sec:intro}

Numerous astronomical observations, such as the rotational curves of galaxies~\cite{Rubin:1970zza,Freeman:1970mx} and the formation of large-scale structures~\cite{Planck:2018vyg}, indicate the existence of an invisible matter component in our Universe, called Dark Matter (DM). Understanding the properties of DM stands as one of the priorities in contemporary physics.
A plausible hypothesis is that DM is constituted by a population of elementary particles that interact with the Standard Model (SM) particles. This hypothesis could be tested by searching for the heat, scintillation light or ionization induced by DM scatterings with nuclei in a dedicated detector~\cite{PandaX-4T:2021bab,LZ:2022lsv,XENON:2023cxc}, the gamma-ray, antimatter particles or neutrinos produced in annihilations in the Milky Way center or other astronomical objects,~\cite{Fermi-LAT:2015att,HESS:2022ygk}, or the detection of unaccounted missing energy in collider experiments~\cite{LHCDarkMatterWorkingGroup:2018ufk}. 

Aside from DM interactions with SM particles, most scenarios allow for DM self-interactions. This possibility has received some attention in the last decade as a possible solution to some tensions between the simplest collisionless cold DM paradigm and observations (see {\it e.g.} \citep{Oman:2015xda,Zentner:2022xux,Boylan-Kolchin:2011lmk,2016A&A...591A..58P,Moore:1994yx,Burkert:1995yz,deBlok:2001rgg,Sand:2002cz}) if DM is strongly self interacting \cite{PhysRevLett.84.3760,Rocha:2012jg,Peter:2012jh,TULIN20181}. While these tensions could also be explained by astrophysical effects, the possibility that DM could have sizable self-interactions is not precluded theoretically, and ought to be tested.

One distinctive prediction of DM self-interactions is the number-changing process that converts $n$ DM particles into $m$ DM particles (\textit{$n  \to  m$ process}). 
For instance, the $2  \to  1$ process corresponds to a semi-annihilation process that leads to a DM particle and a non-DM particle in the final state, which could affect the DM abundance via freeze-out~\cite{Hambye:2008bq,Hambye:2009fg,Arina:2009uq,DEramo:2010keq}. Besides, the $3  \to  2$  and $4\rightarrow 2$ processes have been studied as alternative mechanism to trigger the freeze-out of DM particles from the thermal plasma. This mechanism requires the DM to be a Strongly Interactive Massive Particle (SIMP), and predicts a DM mass of  $\mathcal{O}(10)$ MeV if the freeze-out is induced by the $3\rightarrow 2$ process, and $\mathcal{O}(100)$ keV  if induced by the $4\rightarrow 2$ process
~\cite{Carlson:1992fn,PhysRevLett.113.171301} (see also~\cite{Bernal:2017mqb,Ho:2017fte,Hochberg:2018rjs,Herms:2018ajr, Arcadi:2019oxh,Smirnov:2020zwf}).

The $n\rightarrow m$ processes, although they could play a pivotal role in determining the DM abundance today, are difficult to test, due to the small density of DM particles in galaxies today. A possible exception is the DM spike which is expected to exist around supermassive black holes (SMBHs) at the center of galaxies, and in particular of the Milky Way.  The extremely high DM density in the spike may allow a significant rate for the $n  \to  m$ process, which will result in a ``boosted'' DM particle in the final state. If this process is very efficient, the DM density in the spike will be significantly depleted, which could have implications for interpreting the orbits of S-stars in DM scenarios~\cite{Weinberg:2004nj,Lacroix:2018zmg}. 
Furthermore, due to the high momentum of the final state DM particles, compared to their mass, the DM particles in the final state may induce very characteristic signatures in direct detection and neutrino experiments~\cite{Toma:2021vlw,Aoki:2023tlb,BetancourtKamenetskaia:2025noa}. For instance, the recoil spectrum is expected to differ from other boosted DM scenarios,  such as cosmic-ray (CR) boosted DM~\cite{Yin:2018yjn,Bringmann:2018cvk,Ema:2018bih}, blazar-boosted DM~\cite{Wang:2021jic}, or solar reflection of DM~\cite{Emken:2021lgc}. In this paper, we will perform a systematic study of the role of the  $n  \to  m$ process in the structure and fate of the DM spike around supermassive black holes (SMBHs).

This work is organized as follows: In Sec.~\ref{sec:n_m_processes}, we discuss the main DM processes relevant to self-interactions, including number-changing reactions. In Sec.~\ref{sec:profile-BDM_flux}, we investigate the impact of these processes on the DM spike density profile, accounting for core formation, self-heating, and DM dissolution. In Sec.~\ref{sec:J-factors}, we explore the implications on the galactic $J$-factor and compare with the expectations from collisionless DM. Finally, in Sec.~\ref{sec:conclusions}, we present our conclusions.

\section{DM number changing processes} \label{sec:n_m_processes}

Inside the spike, the DM density can reach very large values, thus allowing to probe interactions of DM particles with SM particles, as well as interactions of DM particles with themselves. In our work we will consider:
\begin{itemize}
\renewcommand{\labelitemi}{$\diamondsuit$}
\item  \textit{Annihilation of $n$ DM particles, producing $m(<n)$ DM particle(s) in the final state} ($n\rightarrow m$ process)

This process (possibly accompanied by SM particles in the final state) modifies the DM density profile, since the ``boosted" DM particles in the final state can transfer energy to the other DM particles in the spike (in a process dubbed  self-heating mechanism~\cite{Chu:2018nki,Kamada:2019wjo}). 
If the heat injection time scale is shorter than the typical age of the DM halo, the density profile is affected by this heat injection and forms a core in the central part. The simplest $n\rightarrow m$ processes that can modify the DM profile are:
\begin{itemize} 
\item[$\blacksquare$] $2\rightarrow 1$: $\chi\chi\rightarrow \chi \, + \,{\rm SM}$

This case requires at least one additional particle in the final state, in order to satisfy the conservation of four-momentum. When the final state contains only one SM particle, with negligible mass compared to the DM particle (for example, $\chi\chi\rightarrow\chi\nu$~\cite{Ma:2007gq,Aoki:2014cja,Ho:2016aye} or $\chi\chi\rightarrow\chi\gamma$~\cite{DEramo:2012fou}), then the 
final state DM particle has an energy $E_\chi  \simeq  5  m_\chi/4$.
\item[$\blacksquare$]  $3\rightarrow 2$: $\chi  \chi  \chi  \to  \chi  \chi$

This process has been discussed in frameworks of SIMP DM. The high density of DM particles in the Early Universe allows a sizable rate for this process that depletes the DM abundance, which continues until the Hubble rate becomes comparable or larger than the $3\rightarrow 2$ rate. 
It has been argued that the observed DM abundance can be reproduced when  ${\cal  O} (100)~\mathrm{MeV}$~\cite{Carlson:1992fn,PhysRevLett.113.171301}. 
In this process, the energy of the final state DM particles is  $E_\chi  \simeq  3  m_\chi /2$.

\item[$\blacksquare$]  $4\rightarrow 2$: $\chi  \chi  \chi \chi  \to  \chi  \chi$

This process has also been discussed in frameworks of SIMP DM, and arises even when the DM particle is self-conjugated, and in particular for sterile neutrinos~\cite{Herms:2018ajr}. The freeze-out of this process was carefully discussed for real scalar DM in~\cite{Arcadi:2019oxh}, and for Majorana fermions in Ref.~\cite{Herms:2018ajr}, leading to the correct DM abundance for DM masses as low as ${\cal  O}(100)~\mathrm{keV}$. Here, the energy of the final state DM particles is $E_\chi  \simeq  2  m_\chi$. 
\end{itemize}
\item  \textit{Annihilation of two DM particles into SM particles only} ($2\rightarrow 0$)

The DM pair annihilation $\chi\chi\rightarrow {\rm SM} \, + \,{\rm SM}$ is generic to many well motivated DM frameworks, and depletes the number of DM particles in the spike, tending to form a plateau in the central regions of our galaxy.  
As is well known, this scenario reproduces the observed DM abundance if 
$\Braket{\sigma_{2\to 0}  v}  \sim  3  \times  10^{-26}~\mathrm{cm}^3~\mathrm{s}^{-1}$ at the epoch of freeze-out. 
On the other hand, if the DM particle transforms non-trivially under an unbroken continuous global or local symmetry, and there is a large asymmetry between DM particle and antiparticle (see {\it e.g.} Ref.~\cite{Kaplan:1991ah}), then the $2\rightarrow 0$ process can be very suppressed in the present Universe.\footnote{Scenarios of asymmetric DM with highly suppressed $2\rightarrow 0$ processes may still allow sizable rates for $n\rightarrow m$ processes, see {\it e.g.} Ref.~\cite{Ghosh:2020lma,Ho:2022tbw}.}
Therefore, we consider both cases, where DM annihilation is efficient and where it is switched off, to examine the impact of this process on the DM density distribution. 
\item  \textit{DM self-scattering process} ($2\rightarrow 2$)

The DM self-scattering $\chi\chi\rightarrow \chi \chi$ transports heat to the exterior of the spike, and tends to flatten the DM distribution in the highest density regions, ultimately leading to an isothermal core where the DM density and velocity dispersion have almost constant values. The $2\rightarrow 2$ process also occurs quite generically in many DM scenarios, either from contact interactions, or from the exchange of light mediators. If the strength of these four-particle interactions is sizable, they may also induce the processes $n\rightarrow m$ with non-negligible rate.
\end{itemize}

The rates of these processes are controlled by the cross-sections $\Braket{\sigma_{n \to  m}  v^{n-1}}$, $\Braket{\sigma_{2\to 0}  v}$ and $\sigma_{2  \to  2}$ for the $n  \to  m$ ($n>m$), $2\to 0$ and $2\to 2$ processes, respectively. On the other hand, rather than working with the cross-sections, it is convenient to use the parametrization proposed in  Ref.~\cite{PhysRevLett.113.171301}. On dimensional grounds, the cross-sections can be expressed in terms of powers of the DM mass, and some effective dimensionless coupling constants that encode all the model dependence of the process. Explicitly,
\begin{eqnarray}
  \Braket{\sigma_{n  \to  m}  v^{n-1}}  
  \!\! &\equiv \displaystyle{
  \frac{ \alpha_{n\rightarrow m}^n}{m_\chi^{3  n  -  4}}}
   &\text{(for $n  \to  m, n>m$)},
   \\
%
%
  \sigma_{2 \to  2} \!\! &\equiv  \displaystyle{ \frac{ \alpha^2_{2\to 2}}{m_\chi^2}}  &\text{(for $2  \to  2$)}. 
  \label{eq:cross-section}
  \end{eqnarray}

One nevertheless expects relations among the effective coupling constants. For instance,  the semi-annihilation $2\to 1$ involves both DM self-interactions as well as portal interactions with the visible sector. Therefore, one generically expects $\alpha_{2\to1} \sim (\alpha_{2\to 2} \alpha_{2\to 0} )^{1/2}$. On  the other hand, in some scenarios the semi-annihilation occurs at the one-loop level, for instance when the SM particle in the final state is a photon. In this case, one expects $\alpha_{2\to 1}\sim (\alpha_{2\to 2}^3 \alpha_{2\to 0})^{1/2}$~\cite{DEramo:2012fou}. In order to understand better the impact of the different contributions, we will consider these three coupling constants as independent parameters. 

\section{DM density profile} \label{sec:profile-BDM_flux}
The DM density profile is directly linked with the possible signatures that can be expected from this astrophysical structure. Namely, large densities translate into large fluxes of particles, whether Standard Model particles from annihilations or boosted DM particles from number-changing self-interactions such as $n\to m$. In particular, the flux of outgoing particles observed at Earth depends on the spatial integral over the density profile, weighted by powers of the DM number density. As a result, small-scale features of the inner spike region can significantly enhance or suppress the expected signal. In what follows, we describe the structure of the DM spike and how it is modified by the different particle interactions discussed in the previous section.

The adiabatic growth of a black hole in the presence of a DM halo is expected to generate a DM overdensity around the black hole, known as a DM spike. The density distribution is mostly modified at distances to the black hole between $4  R_{\rm   sch}$~\cite{Gondolo:1999ef}\footnote{In the relativistic treatment, the inner radius is obtained as $2 R_{\rm  sch}$. In this paper, we take the original value proposed in Ref.~\cite{Gondolo:1999ef} to maintain consistency through the analysis. With this choice, we evaluate the boosted DM flux more conservatively. 
}, where $R_{\rm sch}  =  2  G  M_{\rm  BH}$ is the Schwarzschild radius of the SMBH, and 
the gravitational influence radius, which represents the typical radius where all the particles are bound to the SMBH, and which  is given by $G  M_{\rm BH}/v_0^2$, where $v_0$ is the velocity dispersion of DM particles. 
We will consider for illustration a Milky-Way like galaxy, adopting a velocity dispersion of $v_0=140~\mathrm{km/s}$ near the galactic center, derived from the observed correlation between DM kinematics and the stellar density distribution~\cite{2018A&A...616A..83V}.

Under these assumptions, one can show that 
an initial DM density profile parametrized as $\rho_{\rm  ini}(r)  =  \rho_s  (r/r_s)^{-\gamma}$ develops the following power law behavior~\cite{Gondolo:1999ef}:\footnote{While Ref.~\cite{Gondolo:1999ef} accounted for DM capture effects via a numerical factor, we neglect capture in this work to isolate the impact of $n \to m$ processes on the density and simplify the comparison of DM density profiles.}
\begin{align}
\rho_{\rm  sp}  (r)  =  \rho_R  \left( \frac{R_{\rm sp}}{r} \right)^{\gamma_{\rm  sp}},
  \label{eq:rho_sp}
\end{align}
where $\gamma_{\rm  sp}  \equiv  (9-2  \gamma)/(4  -  \gamma)$ and $R_{\rm  sp}$ is the spike radius, given by
\begin{align}
  R_{\rm  sp}  &=  \alpha_\gamma  r_s  \left( \frac{M_{\rm BH}}{\rho_s  r_s^3} \right)^\frac{1}{3  -  \gamma},
  \label{eq:R_sp}
\end{align}
where the constant $\alpha_\gamma$ must be evaluated numerically, and is $\alpha_\gamma\approx 0.1$ for $\gamma=0.7-1.4$~\cite{Gondolo:1999ef}. The normalization factor $\rho_R$ follows from matching the spike profile to the initial spike at the distance $R_{\rm sp}$, namely $\rho_R  =  \rho_{\rm ini}  (R_{\rm sp})$. In line with our consideration of a Milky-Way like galaxy, we have an initial DM density distribution described by an NFW profile~\cite{Navarro:1995iw,Navarro:1996gj}
\begin{align}
  \rho_{\rm  NFW}  (r)  &= \frac{\rho_s}{\left(\frac{r}{r_s} \right)  \left( 1  +  \frac{r}{r_s} \right)^2},
  \label{eq:NFW}
\end{align}  
with $\rho_s  =  0.184~\mathrm{GeV}/\mathrm{cm}^3$ and $r_s  =  24.42~\mathrm{kpc}$~\cite{Bergstrom:1997fj,Turner:1985si,Bertone:2004pz,Cirelli:2010xx}. The density profile can then be roughly separated into three parts: 
\begin{align}
  \rho_\chi  (r)  =  
  \left\{
  \begin{array}{lll}
  0  &  (r  \leq  4  R_{\rm  sch})  &  \text{: SMBH region}
  \\[2ex]
  \rho_{\rm  sp}  ( r )  &  (4  R_{\rm  sch}  \leq  r  <  R_{\rm  sp})  &  \text{: spike region}
  \\[2ex]
  \rho_{\rm  NFW}  (r)  &  ( r  \geq  R_{\rm  sp} )  &  \text{: NFW region}
  \end{array}
  \right.
  .
  \label{eq:spike_NFW}
\end{align}

The effect of the $n\to m$, $2\to 0$ and $2\to 2$ processes on the structure of the DM spike will be described in detail in the following subsections.

\subsection{Isothermal core formation}
\label{sec:isothermal}

Strong DM self-interactions smooth out a cuspy profile, eventually producing a core~\cite{PhysRevLett.84.3760}.
We estimate that the region of the DM distribution that is affected by the $2\rightarrow 2$ scatterings corresponds to the region of radius $r_1$ where there is at least one scattering during the age of the halo, $t_{\rm age}$~\cite{PhysRevLett.116.041302}. Namely,
\begin{align}
  \frac{4}{\sqrt{\pi}} \cdot  \rho_\chi  (r_1)  \cdot  \frac{\sigma_{2  \to  2}}{m_\chi}  \cdot  v_0  (r_1)  \cdot  t_{\rm  age}  \sim  1. 
  \label{eq:SIDM_r1}
\end{align}
%
Regarding the velocity dispersion $v_0(r)$, we have assumed that the DM particles follow a Maxwell-Boltzmann distribution~\cite{REI65}. In this region the density and velocity distribution are~\cite{PhysRevD.89.023506} (see also \cite{PhysRevD.89.023506} for a numerical simulation):
\begin{align}
  \rho_\chi  (r)  \propto  r^{-\frac{3+k}{4}}, ~~~~
  v  (r)  \propto  r^\frac{1}{2},
  \label{eq:spike_power_law}
\end{align}
where $k$ measures the dependence of the $2\rightarrow 2$ cross-section with the velocity:
  \begin{align}
    \sigma_{2  \to  2}  =  \sigma_0  \left(  \frac{v}{v_0} \right)^{-k}.
  \end{align}
For a contact self-interaction $k=0$, while for a Coulomb-like or long-range self-interaction $k=4$. For concreteness, we focus on the $k=0$ case, which maximizes core formation due to its velocity-independent cross-section. This choice provides the most conservative estimate for flux predictions, as this choice leads to the weakest spike power law index as per Eq.~\eqref{eq:spike_power_law}.

\subsection{Self-heating and core formation}
\label{sec:self-heating}

The final DM particle(s) of the $n  \to  m$ process acquires a kinetic energy of the order of the DM mass, which could be transferred to the DM particles in the spike via $2\rightarrow 2$ scatterings, thus heating it up. The efficiency of the energy transfer within the radial distance $r$ can be parametrized as~\cite{Chu:2018nki,Kamada:2019wjo}, 
\begin{align}
\xi (r)  &=  r  \times  \frac{\rho_\chi(r)  \sigma_{\rm  2  \to  2}}{m_\chi},
\label{eq:capture_rate}
\end{align}
and is defined such that $\xi(r)\leq 1$, the upper limit occurring when all the final DM particles from the $n  \to  m$ process suffer an interaction before reaching the radial coordinate $r$. Note that we are considering only a single scattering before $r$, and that in general in one collision not all the energy of the projectile is transferred to the target. Therefore, even when $\xi(r)=1$, the energy transfer may not be 100\% efficient. However, we will assume for simplicity that all the energy is transferred to the spike when $\xi(r)=1$. This assumption overestimates the efficiency of self-heating and thus the extent to which the spike is depleted. As a result, the expected flux, being based on a more depleted profile, will be conservative.   

To estimate the core radius by the self-heating, 
we compare the typical time scale for the injected heat to dominate the kinetic energy of the whole system. 
The representative heat time can be expressed as~\cite{Kamada:2019wjo}
\begin{align}
  t_{\rm heat}
  =\left(
  \frac{\rho_\chi  (r)}{m_\chi}\right)^{1-n}\frac{1}{\Braket{\sigma_{n  \to  m}  v^{n-1}}}
  \frac{m  v(r)^2}{\xi  \delta  E},
\end{align}
where $\delta  E$ is the injected energy by capturing the final state DM of the $n  \to  m$ process. 
This time scale depends not only on $\Braket{\sigma_{n  \to  m}  v^{n-1}}$ but also on $\sigma_{2  \to  2}$ via the dependence of $\xi$ (Eq.~\eqref{eq:capture_rate}). 
We evaluate the DM density and DM velocity $v(r)$ at the typical radius to characterize the core region, which is determined in the following way:
if the DM density is high enough, $t_{\rm  heat}$ will be smaller than the age of DM density, and the core will be formed. 
We estimate this radius by reading out the radius where $t_{\rm  heat}$ starts to exceed $t_{\rm  age}$ by taking for $\rho_\chi(r)$ the DM density after spike formation from the isothermal core/NFW profile.

Lastly, the high capture rate may directly reduce our boosted DM flux.
We can observe the ratio of $(1  - \xi)$ in the total flux that escapes from the capture process or does not experience scattering to be off from the path. 
We have checked the capture rate is much smaller than $1$ in our interesting parameter region and concluded that this reduction is irrelevant.

\subsection{DM dissolution effects}
\label{sec:dissolution}

Lastly, we consider the modifications of the density distribution by DM number changing processes, namely the $2\rightarrow 0$ and the $n\rightarrow m$ process. The time evolution of the DM number density for the fixed orbit $r$, $n_\chi  (r, t)$, is described by:
\begin{align}
  \dot{n}_\chi  ( r,  t )
  &=
  -  \Braket{\sigma_{2\to 0}  v}    \left( n_{\chi}  (r, t) \right)^2
  -  \frac{n}{n!}  \Braket{ \sigma_{n  \to  m}  v^{n-1} }  \left( n_{\chi}  (r, t) \right)^n, 
  \label{eq:DM_number_dissolution}
\end{align}
where we have assume identical DM particles in the initial state, which results in symmetry factors $1/2$ and $1/n!$ in the annihilation rate for the $2\rightarrow 0$ and $n\rightarrow m$ processes, respectively. Note also that two DM particles disappear in each $2\rightarrow 0$ process, while $n$ DM particles disappear in each $n\rightarrow m$ process. The reason why all DM particles disappear in the $n\to m$ process is because the $m$ particles in the final state are boosted to relativistic velocities. Such particles are able to escape
from their orbit at $r$, which means that, as far as the DM density profile is concerned, all particles are lost.

As an initial condition for Eq.~\eqref{eq:DM_number_dissolution}, we identify $\rho_{\chi}  ( r,  t_{\rm ini} )/m_\chi$ as the DM number density obtained by the procedure detailed in the previous subsections. This treatment is justified because spike formation occurs in a shorter time scale than the age of the DM density~\cite{PhysRevD.89.023506}.
Core formation effects by $2 \to 2$ and $n \to m$ processes coexist with the dissolution effects but all the effects reduce DM density. 
Since our analysis selects the upper bound on the DM density in the central region, the ordering of the process does not matter in the final result.

Suppose only one process dominates the right-hand side of Eq.~\eqref{eq:DM_number_dissolution}, and let us call this process \textit{$N  \to  M$ process}.
In our setup, the $N  \to  M$ process can be either the DM annihilation or the $n  \to  m$ process.
Then, we can derive an analytical solution by integrating Eq.~\eqref{eq:DM_number_dissolution} from $t_{\rm ini}$ to $t$ ($t_{\rm age} = t  -  t_{\rm  ini}$). 
\begin{align}
  \rho_\chi  (r,t)  
  &=  
  \frac{\rho_{\rm  pl}^{N  \to  M}  \rho_\chi  (r,  t_{\rm  ini})}{\left[ ( \rho_{\rm  pl}^{N  \to  M} )^{N-1}  +  \rho_\chi  ( r,  t_{\rm  ini} )^{N-1}  \right]^{\frac{1}{N-1}}},
  \label{eq:rho_chi}
\end{align}
where we have introduced the ``plateau density" around the central region: 
\begin{align}\label{eq:plat_dens}
  \rho_{\rm  pl}^{N  \to  M}
  &\equiv
  m_\chi  \left( \frac{(N-2)!}{\Braket{\sigma_{N  \to  M}  v^{N-1}}  t_{\rm age}} \right)^{\frac{1}{N  -  1}}.
\end{align}
As seen from Eqs.~\eqref{eq:DM_number_dissolution}-\eqref{eq:plat_dens}, only the initial amount of DM particles ($N$) is relevant for the dissolution effect.  Therefore, for concrete processes, we obtain
%

\begin{align}
  \rho_{\rm  pl}^{2  \to  1}  
  &\equiv  \frac{m_\chi}{\Braket{\sigma_{2  \to  1}  v}  t_{\rm age}} 
  \nonumber
  \\
  &=  2.7  \times  10^{-1}~
  \mathrm{GeV} \mathrm{cm}^{-3}
  \left(\frac{\alpha_{2\to1}}{1.0}\right)^{-2}\left( \frac{m_\chi}{1~\mathrm{GeV}} \right)^3
  \left( \frac{t_{\rm age}}{10~\mathrm{Gyr}} \right)^{-1},
  \label{eq:rho_pl_2toM}
  \\
  \rho_{\rm  pl}^{3  \to  2}  
  &\equiv  \frac{m_\chi}{\sqrt{\Braket{\sigma_{3  \to  2}  v^2}  t_{\rm age}}} \nonumber
  \\
  &=  1.9  \times  10^{13}~
  \mathrm{GeV} \mathrm{cm}^{-3} \left(\frac{\alpha_{3\to2}}{1.0}\right)^{-\frac32}
  \left( \frac{m_\chi}{10~\mathrm{MeV}} \right)^{\frac72}
  \left( \frac{t_{\rm age}}{10~\mathrm{Gyr}} \right)^{-  \frac{1}{2}},
  \\
  \rho_{\rm  pl}^{4  \to  2}  
  &\equiv  m_\chi  \left( \frac{2}{ \Braket{\sigma_{4  \to  2}  v^3}  t_{\rm age}} \right)^{\frac{1}{3}} \nonumber
  \\
  &=  4.5  \times  10^{12}~
  \mathrm{GeV} \mathrm{cm}^{-3} \left(\frac{\alpha_{4\to2}}{1.0}\right)^{-\frac43}
  \left( \frac{m_\chi}{100~\mathrm{keV}} \right)^{\frac{11}{3}} 
  \left( \frac{t_{\rm age}}{10~\mathrm{Gyr}} \right)^{-  \frac{1}{3}}.
\end{align}
The value of $\rho_{\rm  pl}$ is the upper bound on DM density for given $(r,t)$.
The reference values for $\alpha_{\rm eff}$ and $m_\chi$ lead to the correct DM abundance via freeze-out for the $3\to2$ and $4\to2$ process, respectively \citep{PhysRevLett.113.171301}.
To treat the general situation where more than one DM number changing process contributes, we solve Eq.~\eqref{eq:DM_number_dissolution} numerically to determine the current DM density.
As in the case of the isothermal core formation and the self-heating core, where we found the critical radii that determine the boundary where each effect is relevant, we also define a ``dissolution'' radius $R_{\rm diss}$. It denotes the radius of transition between the initial density profile before dissolution and the upper bound given by $\rho_{\rm pl}$.

As mentioned previously, to illustrate the impact of the various DM number conserving and violating processes on the structure of the DM spike we will consider a DM density distribution described by a NFW profile, as in Eq.~\eqref{eq:NFW}. Furthermore, the Milky Way hosts in its center a SMBH with an estimated mass of $M_{\rm BH}  \simeq  4.15  \times  10^6~M_\odot$~\cite{Gravity:2019nxk}, which produces a spike with size $R_{\rm  sp}  \sim  18.7~\mathrm{pc}$, and density profile with $\gamma_{\rm sp}=7/3$ and normalization $\rho_R\sim240.3~\mathrm{GeV}/\mathrm{cm}^3$. It has been argued that the spike in the Milky Way may be depleted by interactions with stars, since the estimated time scale of stellar heating is a factor of 10 smaller than the age of the central SMBH, Sgr A$^*$~\cite{Sandick:2016zeg}. This heating might not occur in other galaxies, which is the main focus of our work, but we keep the Milky-Way parameters as an archetype of DM halo with a supermassive black halo at its center to investigate the impact of particle physics processes in determining the fate of the spike.

The impact of these processes on the density distribution inside the spike depends on the DM mass, and the cross-sections for the processes $2\to 0$, $2\to 2$ and $n \to m$.
To analyze this large parameter space we will consider eight benchmark points, corresponding to different values for the cross-section of the various DM reactions, fixing for concreteness $m_\chi=100$ MeV. In particular, we focus on two representative $n\to m$ processes: the $2\to1$ semi-annihilation and the $3\to2$ interaction, the latter serving as a representative for scenarios with $n\geq3$. The parameters are listed in Table~\ref{table:BP}. 

In benchmark points A-D, we fix the cross-section for the $3\to 2$ process to zero and the cross-section for the $2\to 1$ process to $\langle \sigma v\rangle_{2\to 1}=10^{-26}\,{\rm cm}^3/{\rm s}$, whereas in  benchmark points E-H, we fix the cross-section for the $2\to 1$ process to zero and the cross-section for the $3\to 2$ process to $\langle \sigma v^2\rangle_{3\to 2}=10^{-57}\,{\rm cm}^6/{\rm s}$. In all benchmark points we set the cross-section for the $2\to 0$ process to zero, so that there is only one DM number changing process at a time. The $2\to 1$ and $3 \to 2$ not only deplete the number of DM particles in the spike, but also produce a component of boosted DM, that can heat up the spike in the presence of DM self-interactions. We therefore consider for each set of benchmark points different values for cross-section for the $2\to 2$ scatterings, which transport heat to the center of the spike, from very small interaction rate to large interaction rate. For each benchmark point, we also show the values of the respective coupling constants.

\begin{table}[t]
\renewcommand{\arraystretch}{1.5}
\centering
\resizebox{\textwidth}{!}{
\begin{tabular}{|c||c|c|c||c|c|c|}
\hline
Benchmark  & $\langle \sigma v\rangle_{2\rightarrow 1}$ [$\frac{\mathrm{cm}^3}{\mathrm{s}}$] & $\frac{\sigma_{2\rightarrow 2}}{m_\chi}$ [$\frac{\mathrm{cm}^2}{\mathrm{g}}$] & $\langle \sigma v^2\rangle_{3\rightarrow 2}$ [$\frac{\mathrm{cm}^6}{\mathrm{s}}$]& $\alpha_{2\to 2}$ & $\alpha_{2\to 1}$& $\alpha_{3\to 2}$ \\
\hline
A  &  $10^{-26}$ &$10^{-10}$ &0 &$2.14\times10^{-5}$ &$2.93\times10^{-6}$ & 0\\
B & $10^{-26}$ &$5\times10^{-8}$ &0&$4.78\times10^{-4}$ &$2.93\times10^{-6}$ & 0 \\
C &  $10^{-26}$ &$5\times10^{-5}$ & 0 &$1.51\times10^{-2}$ &$2.93\times10^{-6}$ & 0\\
D &  $10^{-26}$ &$5\times10^{-2}$ & 0 &$4.78\times10^{-1}$ &$2.93\times10^{-6}$ & 0\\
\hline
E &  $0$ &$10^{-7}$ &$10^{-57}$  &$6.77\times10^{-4}$ &0 & $4.81\times10^{-2}$\\
F&  $0$ &$5\times10^{-5}$ &$10^{-57}$   &$1.51\times10^{-2}$ &0 & $4.81\times10^{-2}$\\
G  &  $0$ &$10^{-3}$ &$10^{-57}$ &$6.77\times10^{-2}$ &0 & $4.81\times10^{-2}$ \\
H &0 &$5\times10^{-2}$ & $10^{-57}$ & $4.78\times10^{-1}$ &0 & $4.81\times10^{-2}$ \\
\hline
\end{tabular}}
\caption{Parameters of the profiles showcased in Figs.~\ref{fig:profiles-AD} and~\ref{fig:profiles-EH}.}
\label{table:BP}
\end{table}

\begin{figure}[t!]
\centering
\includegraphics[height=0.4\textwidth]{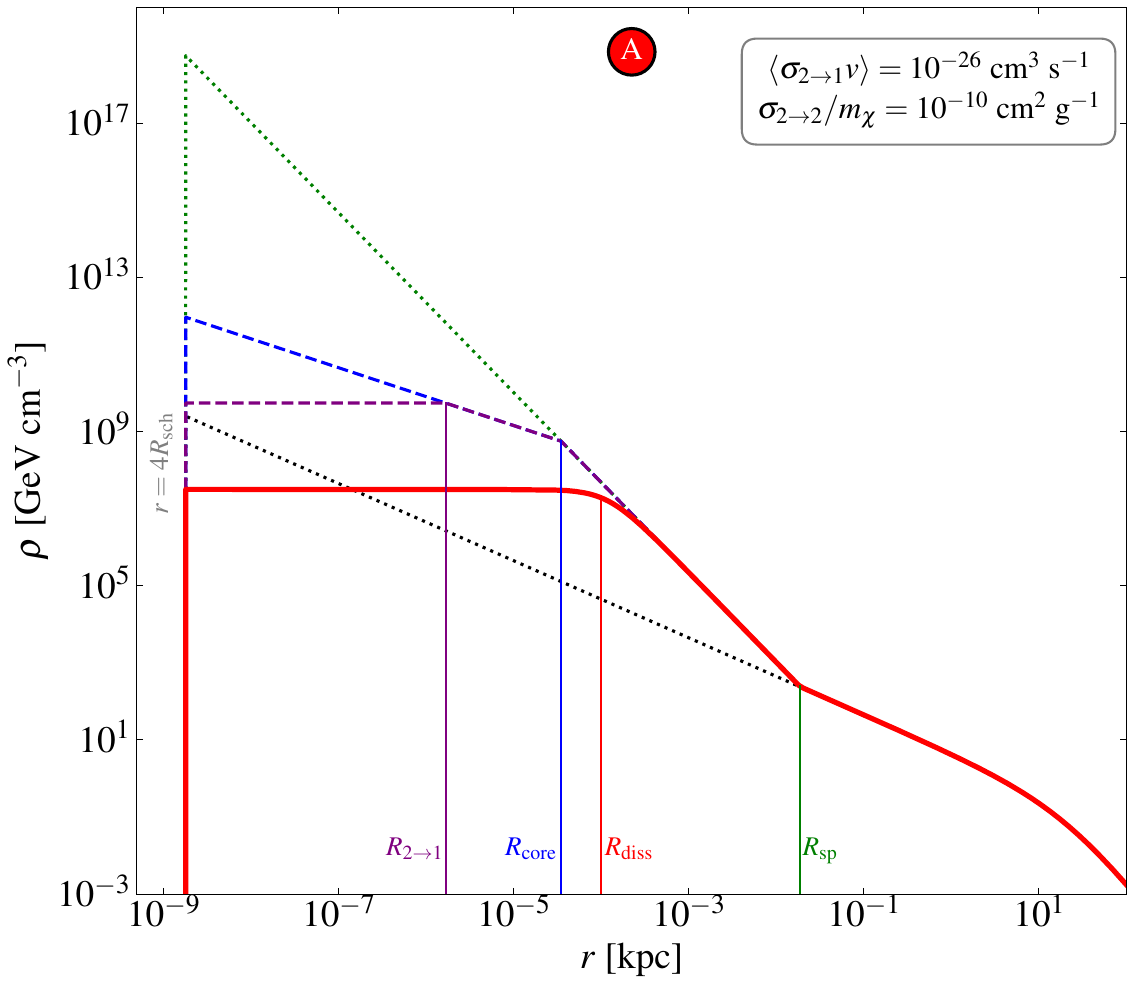}
\includegraphics[height=0.4\textwidth]{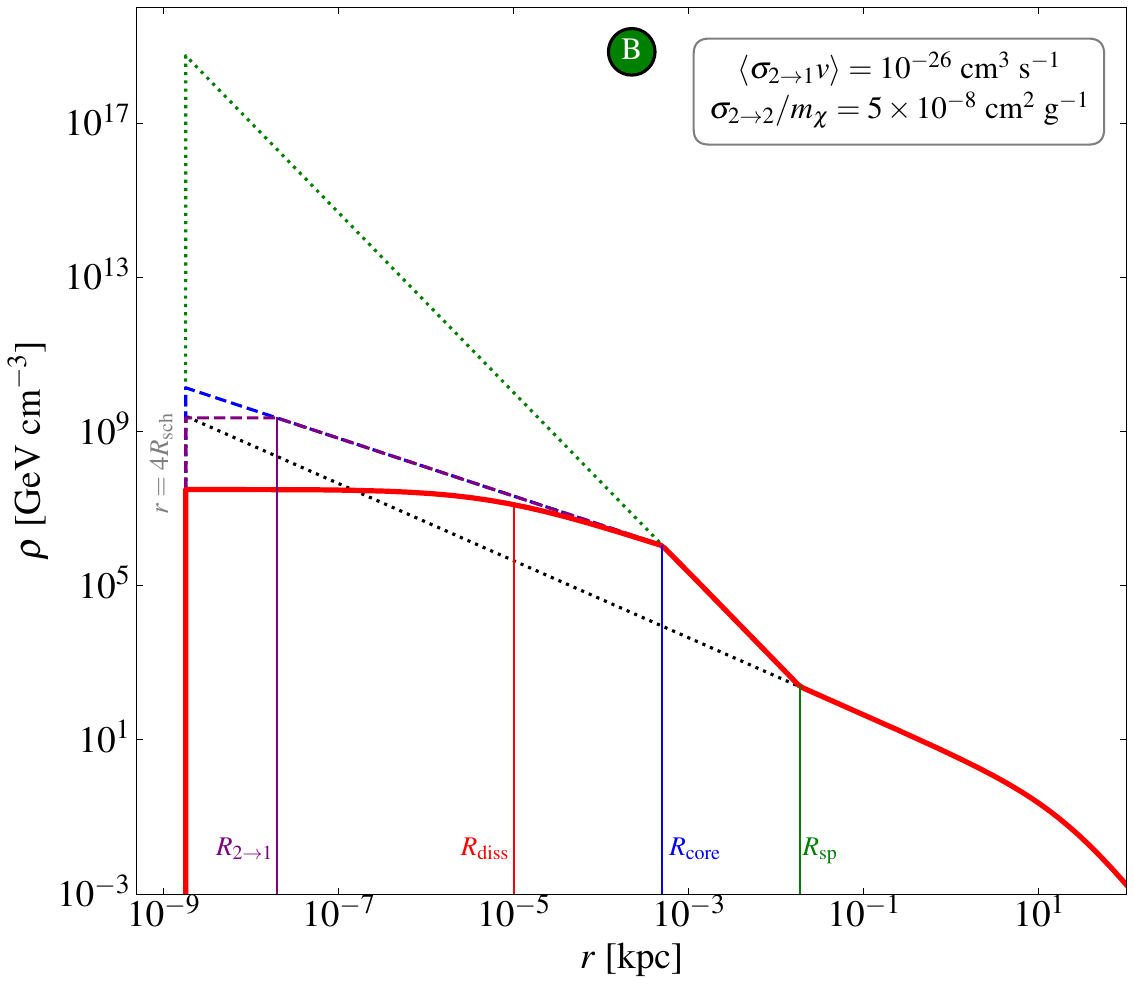}\\
\includegraphics[height=0.4\textwidth]{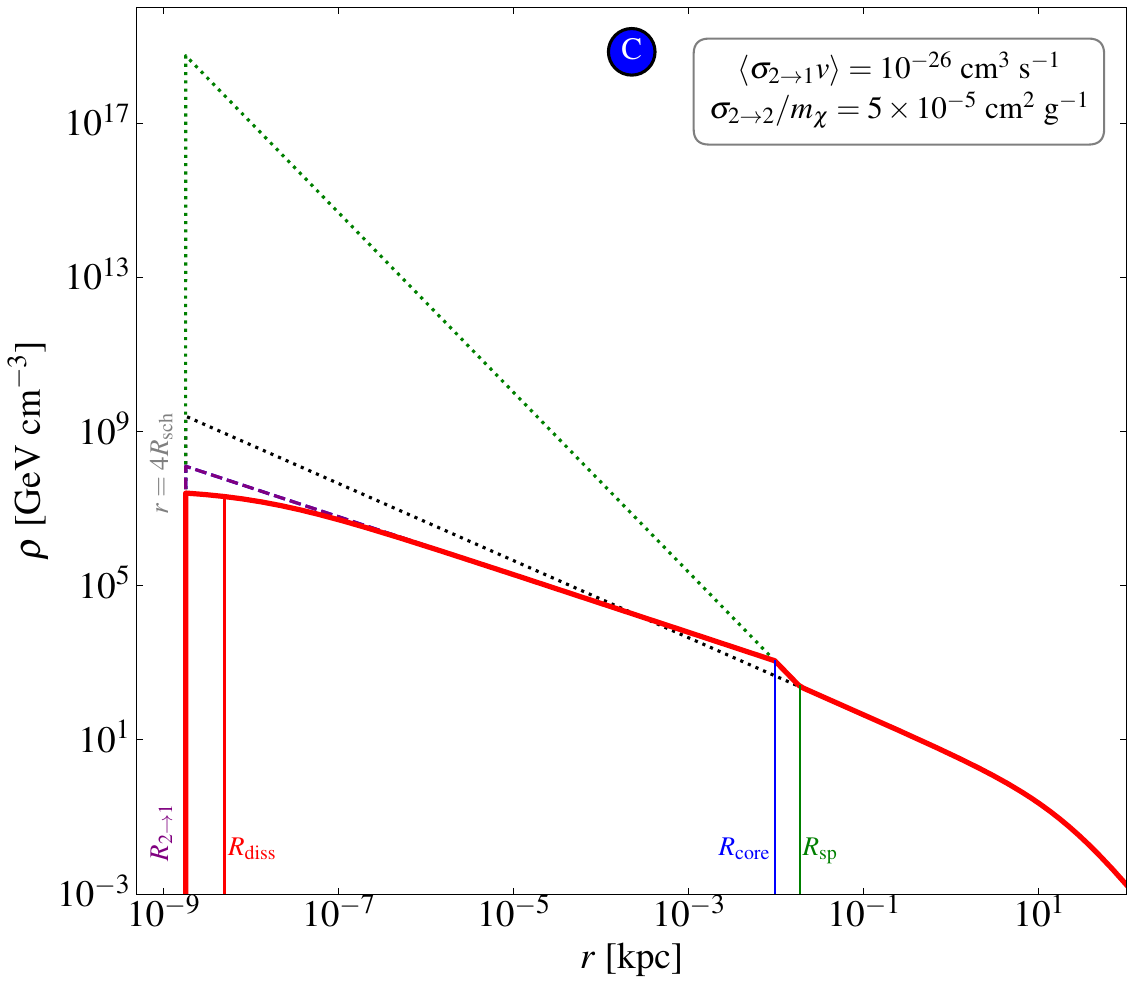}
\includegraphics[height=0.4\textwidth]{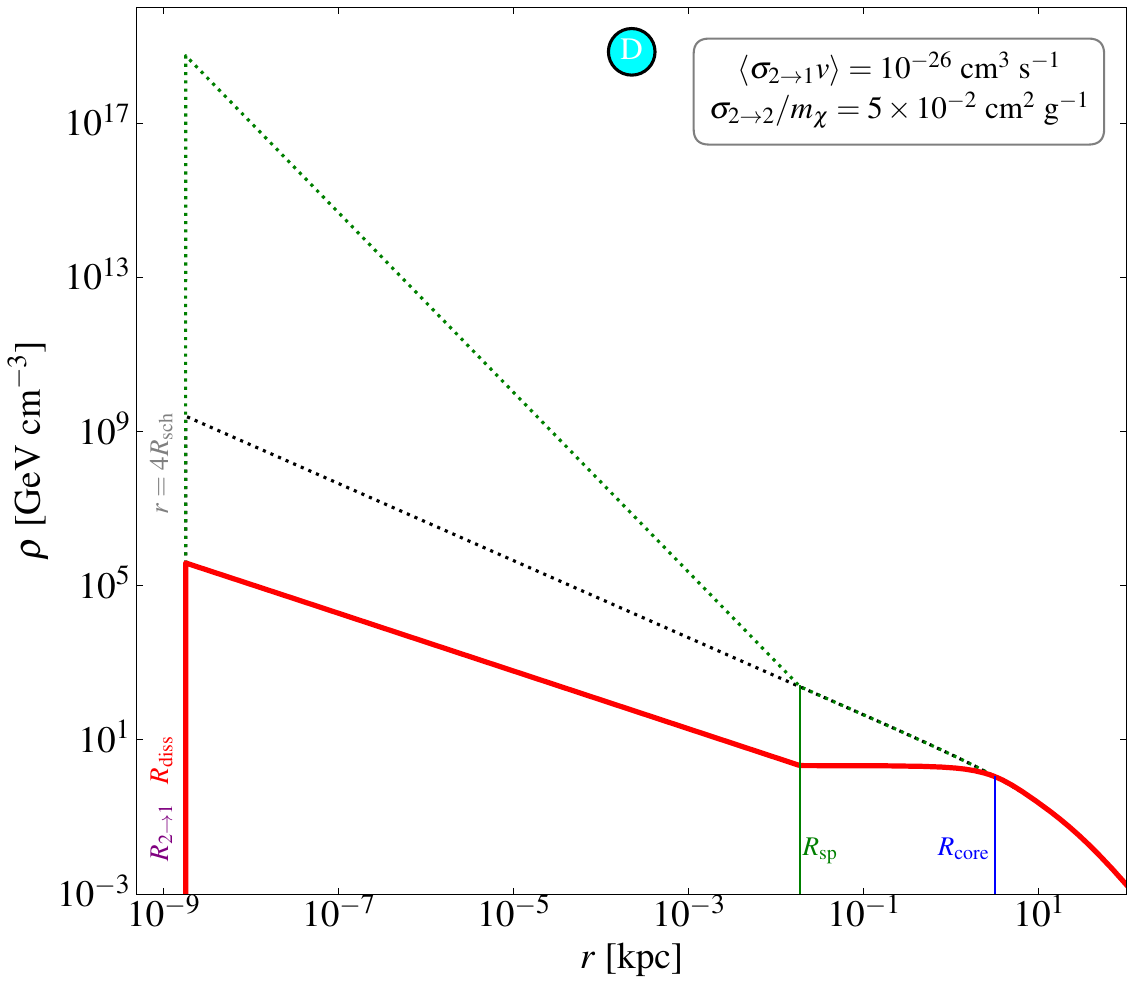}
        \caption{DM halo profiles for $\langle\sigma_{2\to 1}v\rangle=10^{-26}\,{\rm cm}^3\,{\rm s}^{-1}$, $\langle\sigma_{2\to 0}v\rangle=0$, $\langle\sigma_{3\to 2}v^2\rangle=0$, and $\sigma_{2\to 2}/m_\chi=10^{-10},~5\times 10^{-8},~ 5\times 10^{-5}$ and  $5\times 10^{-2}\,{\rm cm}^2\,{\rm g}^{-1}$ for the top-left, top-right, bottom-left and bottom-right panel, respectively. The black dotted line indicates the NFW profile, the green dotted line the spike profile for cold DM ($\propto r^{-7/3}$), the blue dashed line the effect on the spike due to the isothermal core formation only (see Sec.~\ref{sec:isothermal}), the purple dashed line the effect of the $2\to 1$ processes via self-heating (see Sec.~\ref{sec:self-heating}), and the red line the combined effect of all processes including the dissolution (see Sec.~\ref{sec:dissolution}).}  
	\label{fig:profiles-AD}
\end{figure}

\begin{figure}[t!]
\centering
\includegraphics[height=0.4\textwidth]{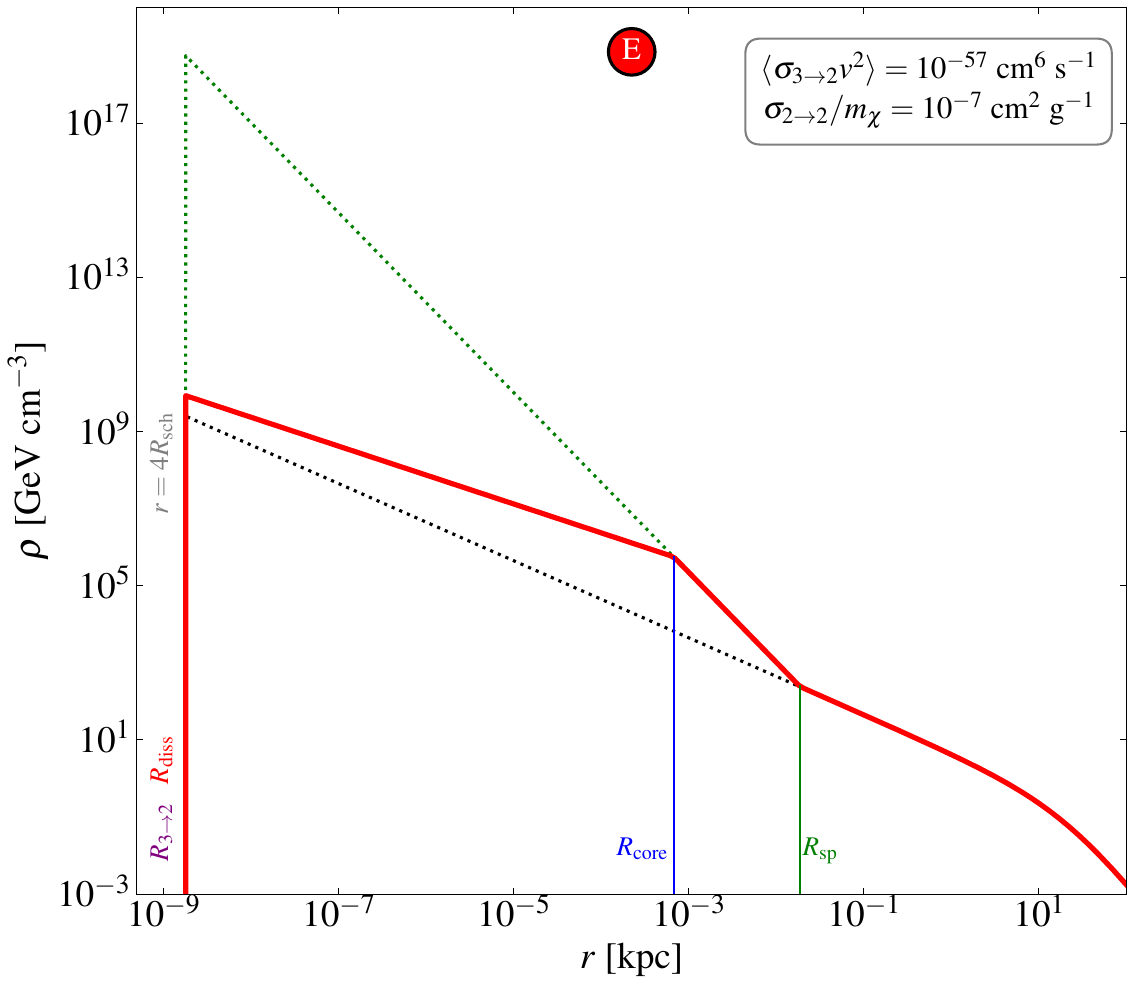}
\includegraphics[height=0.4\textwidth]{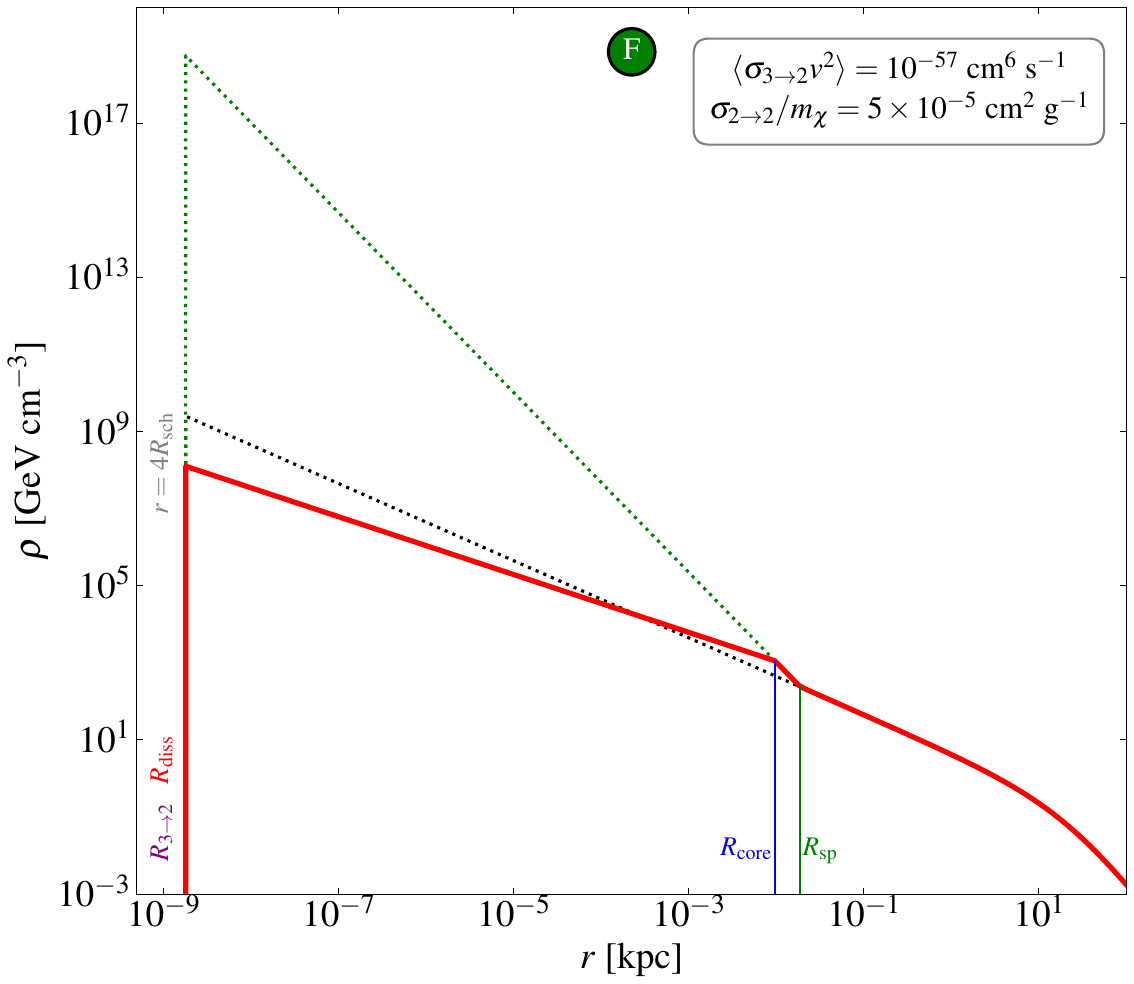}\\
\includegraphics[height=0.4\textwidth]{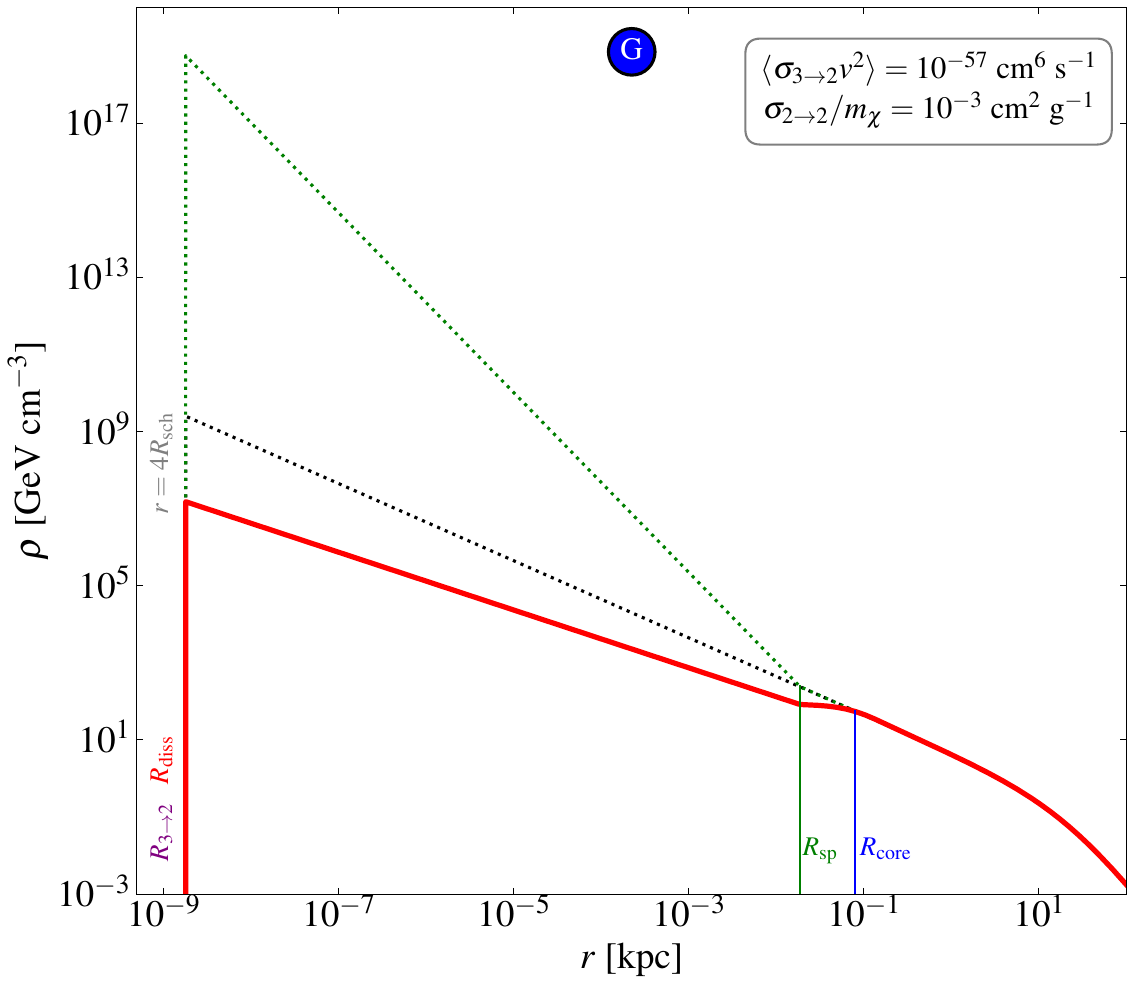}
\includegraphics[height=0.4\textwidth]{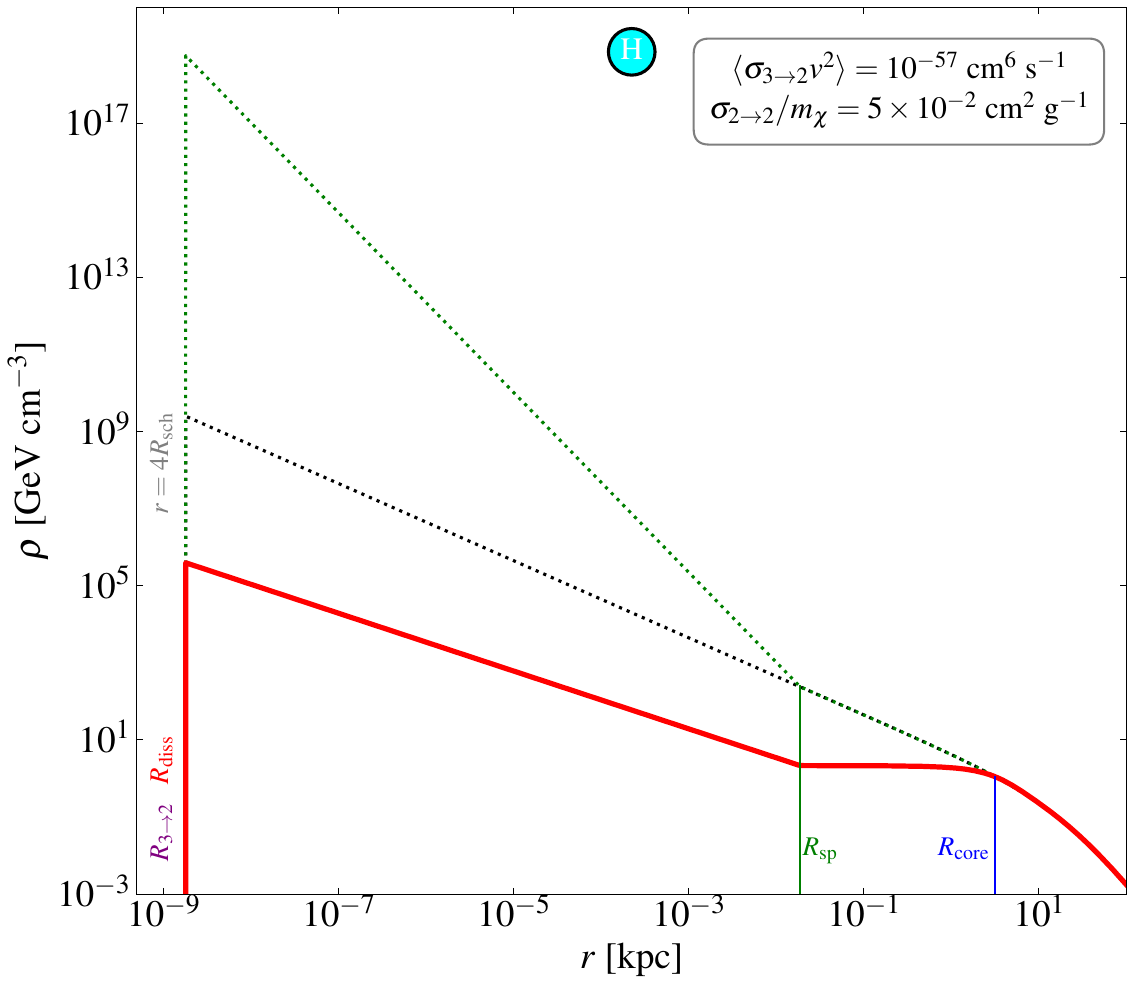}
        \caption{Same as Fig.~\ref{fig:profiles-AD}, but for $\langle\sigma_{2\to 1}v\rangle=0$, $\langle \sigma_{3\to 2}  v^2\rangle=10^{-57}\,{\rm cm}^6\,{\rm s}^{-1}$ and $\sigma_{2\to 2}/m_\chi=10^{-7},~5\times 10^{-5},~ 10^{-3}$ and  $5\times 10^{-2}\,{\rm cm}^2\,{\rm g}^{-1}$ for the top-left, top-right, bottom-left and bottom-right panel, respectively.}
	\label{fig:profiles-EH}
\end{figure}

The resulting profiles are shown in Fig.~\ref{fig:profiles-AD} for profiles A-D, and Fig.~\ref{fig:profiles-EH} for profiles E-H. The dotted black, dotted green, blue dashed, purple dot-dashed and the solid red curve represent, respectively, the standard NFW profile, the spike enhancement, the modified spike profile from the isothermal core,  the core profile due to the $n  \to  m$ self-heating process, and the DM profile expected at the current time after being dissolved by DM number changing processes. 
The vertical grid lines indicate the location of the  spike radius ($R_{\rm  sp}$ defined in Eq.~\eqref{eq:R_sp}), 
the  isothermal core radius ($R_{\rm c}$, evaluated as $r_1$), the self-heating core radius ($R_{n  \to  m}$), and the dissolution radius ($R_{\rm  diss}$). 

Both from Fig.~\ref{fig:profiles-AD}  and Fig.~\ref{fig:profiles-EH} , we note that for small values of $\sigma_{2\to 2}/m_\chi$ (top left panels), the effect of the heating by the self-interactions is negligible. For the $2  \to  1$ process, the spike profile is only affected by the disappearance of DM particles via dissolution (in this regime $R_{\rm sp}\gg R_{\rm diss}\gg R_{\rm core}, R_{2\to1}$). As the self-interaction cross-section increases (top right and bottom left plots), the effects of the core formation and the heating become more and more important. As a result, the spike becomes less cuspy and the dissolution processes are only significant in the innermost parts of the spike (in this regime,  $R_{\rm sp}\gg R_{\rm core}\gg R_{\rm diss}\gg R_{2\to 1}$). Finally, for large values of the self-interaction cross-section, a core is formed even at large distances from the black hole, seeding the formation of a spike with a density profile much less pronounced than in the purely collisionless case (in this regime,  $R_{\rm core} \gg R_{\rm sp}\gg R_{\rm diss}\gg R_{2\to 1}$).

For the $3\to2$ process, we observe a similar qualitative behavior. However, in the benchmark scenarios we consider, neither the self-heating nor the dissolution induced by the $3\to2$ interactions significantly alter the DM density profile. Even when the $3\to2$ cross-section is increased, the shape of the profile remains largely determined by isothermal core formation. This is because the rate of energy injection and particle depletion from $3\to2$ processes remains subdominant compared to the effects of $2\to2$ interactions across most of the parameter space. We have verified that this behavior remains for general $n \to m$ processes with $n \geq 3$, where the high cross-sections required to turn depletion or self-heating dominant correspond to couplings that are no longer perturbative. Furthermore, we have confirmed that this is a general statement for $n\to m$ processes with $n\geq3$, which is why we have chosen the $3\to2$ case as a representative.

Finally, we note that throughout this analysis we have set $\langle \sigma_{2 \to 0} v \rangle = 0$. This choice isolates the effects of self-interactions and number-changing processes involving DM in the final state. Including the $2 \to 0$ annihilation process would lead to additional depletion of the DM density via dissolution, but unlike the $2 \to 1$ or $3 \to 2$ processes, it does not induce isothermal or self-heating core formation.

\clearpage

\section{Implications for the $J$-factors} \label{sec:J-factors}

The annihilation of DM particles in spikes around supermassive black holes could lead to observable signatures at Earth, either from the SM particles produced in the $2\rightarrow 0$ or $2\rightarrow 1$ annihilations, or from the boosted DM particles produced in the $n\rightarrow m$ processes, with $n>m>0$. Given that the spike subtends a very small solid angle as seen from the Earth, this can be regarded as a point source. If the SM particles in the $2\rightarrow m$ are photons ({\it i.e.} the annihilation $\chi\chi\rightarrow \gamma\gamma$ or the semi-annihilation $\chi\chi\rightarrow\chi\gamma$), the flux of photons at Earth can be approximated as
\begin{align}\label{eq:Phi_BDM_2}
  \Phi_{\gamma}
  =  \frac{1}{4\pi d^2}
  \frac{2-m}{2}  J^{\rm spike}_2  \Braket{\sigma_{2  \to  m}  v},
\end{align}
while the flux of boosted DM particles produced in the $n\rightarrow m$ process can be approximated by:
\begin{align}\label{eq:Phi_BDM_n}
  \Phi_{\rm BDM}
  =  \frac{1}{4\pi d^2}
  \frac{m}{n!}  J^{\rm spike}_n  \Braket{\sigma_{n  \to  m}  v^{n-1}},
\end{align}
where $d$ is the distance from the source to the detector, and we have defined a generalized $J$-factor for a process with $n$ initial DM particles as:
\begin{align}
  J^{\rm spike}_n  \equiv  4  \pi  \int_0^{R_{\rm sp}}  dr~r^2 \big(\rho_\chi(r)/m_\chi\big)^n.
  \label{eq:Jnfactor}
\end{align}
It is important to note that the density profile itself is affected by the $n\rightarrow m$ processes as discussed in Sec.~\ref{sec:profile-BDM_flux}, and therefore the flux is not a simple  linear function of $\Braket{\sigma_{n  \to  m}  v^{n-1}}$. 

To investigate the impact of the various $n\rightarrow m$ processes on the flux of DM particles (or photons) from a spike at the center of a galaxy, we will compare the $J_n$ factors obtained from Eq.~(\ref{eq:Jnfactor}) with the $J_n$ factors calculated from the baseline NFW profile, neglecting the possible formation of a spike, and neglecting any effect from $n\rightarrow m$ processes or stellar heating, namely
\begin{align}
  J^{\rm NFW}_n  \equiv  4  \pi  \int_0^{R_{\rm sp}}dr~  r^2 \big(\rho_{\rm NFW}(r)/m_\chi\big)^n
\end{align}
with $\rho_{\rm NFW}(r)$ given in Eq.~(\ref{eq:NFW}).

\begin{figure}[tb]
	\centering
	\includegraphics[width=0.45\textwidth]{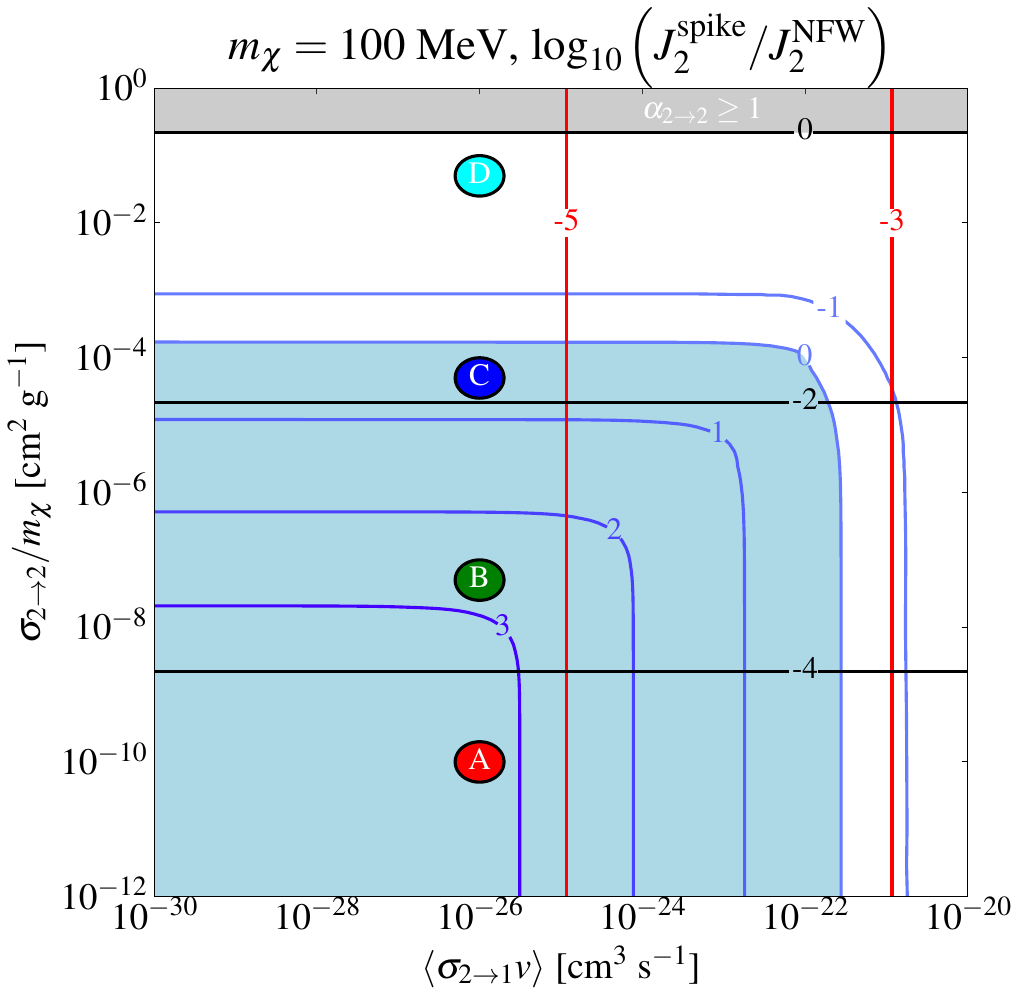}
	\includegraphics[width=0.45\textwidth]{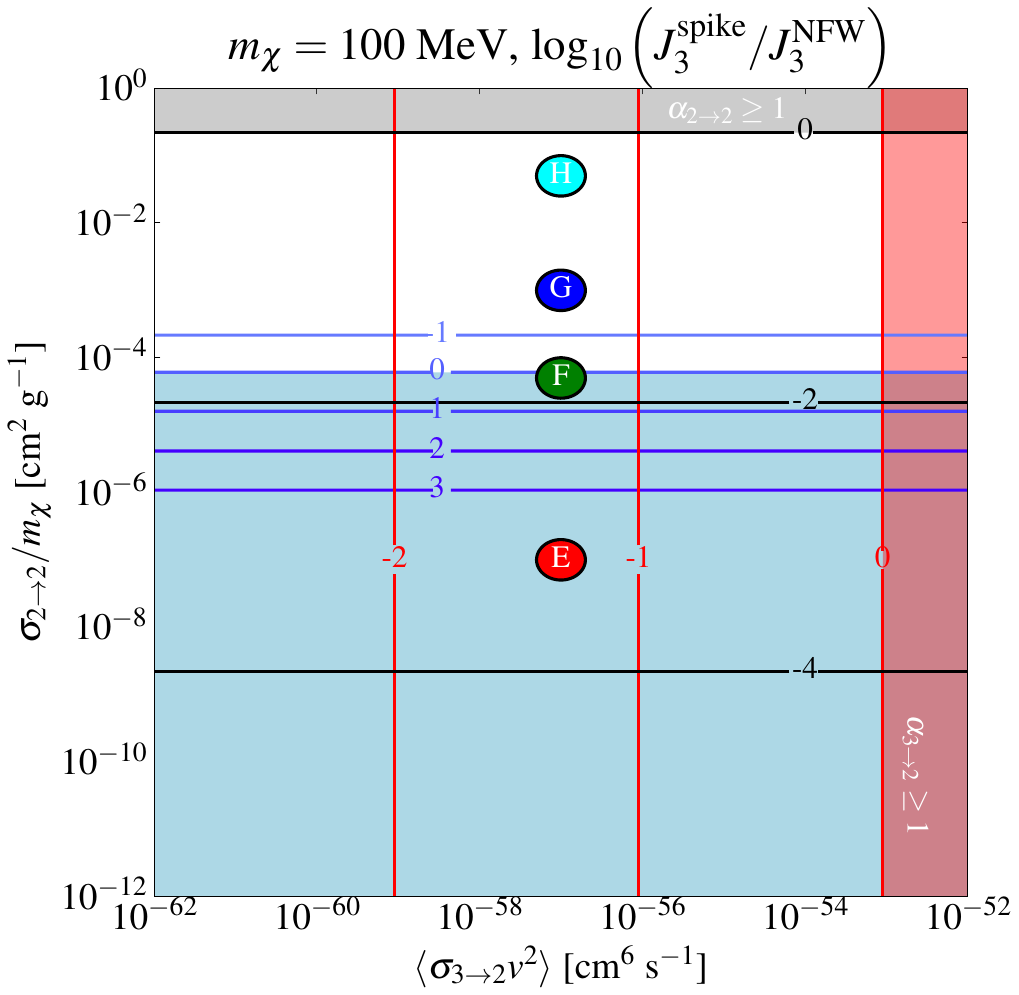}
	\caption{
	Contours of the ratios of $J_2$ (left panel) and $J_3$ (right panel) relative to the naive expectation from the NFW profile (blue lines) due to the effect of the self-interactions and the number changing processes $2\to 1$ (left panel) and $3\to 2$ (right panel), while keeping the rates of all other processes to zero. The plot also shows contours of constant $\alpha_{2\to 2}$ (black lines) and constant $\alpha_{2\to 1}$ or $\alpha_{3\to 2}$ (red lines). In both plots, we fixed $m_\chi=100$ MeV.
	}
	\label{fig:mchi-100MeV_2to1-(1110)}
\end{figure}

In Fig.~\ref{fig:mchi-100MeV_2to1-(1110)}, left panel we show  as blue lines the contours of the ratios $J_2^{\rm spike}/J_2^{\rm NFW}$ in the plane $\langle \sigma_{2\to 1} v\rangle$ vs. $\sigma_{2\to 2}/m_\chi$, when the $3\to 2$ process has a vanishing rate, and when $m_\chi=100$ MeV. We also highlight the benchmark points defined in 
Table \ref{table:BP}. We also show for reference the values of the couplings constants $\alpha_{2\to 2}$ (black) and $\alpha_{2\to 1}$ (red), as well as the region where these couplings become no longer perturbative. We find that the $J_2$ factor can be enhanced compared to the ``naive" expectation from the NFW profile if $\sigma_{2\to 2}/m_\chi\lesssim 10^{-4}\,{\rm cm}^2\,{\rm g}^{-1}$ and when $\langle \sigma_{2\to 1}\rangle \lesssim 10^{-22} \,{\rm cm^2}\,{\rm s}^{-1}$. Otherwise, the $J_2$ factor is smaller than the one predicted from the NFW profile, with implications for the observation of photon or neutrino signals from galaxies, or from the boosted DM signals. 

Finally, we show in the right plot the ratio $J_3^{\rm spike}/J_3^{\rm NFW}$ in the plane $\langle \sigma_{3\to 2} v\rangle$ vs. $\sigma_{2\to 2}/m_\chi$ when the $2\to 1$ process has a vanishing rate. We also highlight the benchmark points defined in 
Table \ref{table:BP}. In this case, the region where the $J_3$ factor is larger than the naive expectation from the NFW profile is much more extended, and covers all the values of $\langle\sigma_{3\to 2}v^2\rangle$ compatible with the perturbative condition. On the other hand, when $\sigma_{2\to 2}/m_\chi\gtrsim 10^{-4}\,{\rm cm}^2\,{\rm g}^{-1}$, the effects of the heating and the generation of a isothermal core become significant and $J_3$ is lowered compared to the result expected from the NFW profile.

\section{Conclusions}
\label{sec:conclusions}
The DM self-interaction is one of the most fundamental parameters for characterizing particle DM properties, and is crucial for correctly determining the DM density distribution in the innermost parts of a galaxy. From a particle physics viewpoint, the DM self-interaction generically leads to processes with $n$ DM particles in the initial state and $m(<n)$ DM particles in the final state, possibly accompanied by SM particles. In this work, we have presented a comprehensive analysis of the implications of $n\rightarrow m$ processes,  $n>m\geq 0$, for the density profile of the DM spike which is expected to surround the supermassive black hole at galactic centers.

We have considered the isothermal core formation due to $2\rightarrow 2$ scatterings. Furthermore, due to the high density of DM particles in the spike, the rate for $n\rightarrow m$ processes with $n>2$ could have a non-negligible rate. Concretely, we have included the self-heating of the spike due to $3\rightarrow 2$ (or $4\rightarrow 2$) processes and the subsequent core formation due to $2\rightarrow 2$ scatterings, as well as the dissolution effects due to $n\rightarrow m$ processes ($n>m\geq 0$). In our work we have focused on the impact of the particle physics processes on the spike, which should occur in all spikes, and we have neglected possible astrophysical effects, such as the stellar heating, which depends on the characteristics of each galaxy. 

We have analyzed in detail the impact of all these effects for some representative points of the parameter space, and we have found that the all these effects make a significant impact in the DM distribution in the innermost parts of a galaxy, and ought not to be neglected in phenomenological applications. 

\section*{Acknowledgments}
\noindent
We thank Ayuki Kamada for useful discussions on the core formation of the DM density profile. 
MF acknowledges the Mainz Institute for Theoretical Physics (MITP) of the Cluster of Excellence PRISMA$^+$ (Project ID 390831469) for feedback at the last stage of this collaboration during ``The Dark Matter Landscape: From Feeble to Strong Interactions''. 
The work of BBK, MF, and AI is supported by the Collaborative Research Center SFB1258 and by the Deutsche Forschungsgemeinschaft (DFG, German Research Foundation) under Germany's Excellence Strategy - EXC-2094 - 390783311. 
The work of TT is supported by JSPS KAKENHI Grant Number 25H02179 and 25K07279.

\bibliographystyle{JHEP}
\bibliography{bibliography}

\end{document}